\documentclass[11pt,a4paper]{article}
\usepackage[T1]{fontenc}
\usepackage{lmodern}
\usepackage[margin=20mm]{geometry}
\usepackage{amsmath,amssymb,graphicx}
\usepackage[table]{xcolor}
\usepackage[hidelinks]{hyperref}
\usepackage{cite}

\title{Confocal ellipsoidal displaced Gregorian structure for stand-off millimeter-wave imaging}
\author{Mohammad Hossein Koohi Ghamsari and Mehdi Ahmadi-Boroujeni\\
\small Department of Electrical Engineering, Sharif University of Technology, Tehran, Iran\\
\small Corresponding author: mohammad.koohi@sharif.edu}
\date{}

\begin{document}
\maketitle

\noindent\textbf{Author-accepted manuscript.} Published in \emph{Journal of the Optical Society of America B}, \textbf{42}(6), 1298--1309 (2025). \href{https://doi.org/10.1364/JOSAB.561816}{doi:10.1364/JOSAB.561816}.

\medskip
\noindent{\footnotesize \textcopyright\ 2025 Optica Publishing Group. One print or electronic copy may be made for personal use only. Systematic reproduction and distribution, duplication of any material in this paper for a fee or for commercial purposes, or modifications of the content of this paper are prohibited. \href{https://opg.optica.org/josab/abstract.cfm?uri=josab-42-6-1298}{Online journal abstract.}}

\begin{abstract}
Reflective-based structures are favorable candidates as the optical section of millimeter-wave (mm-wave) and terahertz (THz) stand-off imaging systems due to decreasing frequency-dependent losses and aberrations compared to refractive-based counterparts. In this paper, we propose a simple confocal ellipsoidal displaced Gregorian (CEDG) dual-mirror configuration for stand-off imaging systems that exhibits superior performance in terms of mitigating optical blockage and enabling object scanning. The proposed structure is composed of a standard elliptical main mirror (MM), which focuses the emanated rays on the focal plane, and a modified secondary mirror (SM) that reduces the optical blockage effect and improves the imaging performance of the structure. A detailed step-by-step procedure is proposed for constructing the structure, followed by an analytical study and deriving closed-form design equations. By performing ray tracing simulations, it is observed that the optical blockage of the proposed CEDG structure is reduced compared with the standard Gregorian. Also, the resolution and the depth-of-focus (DoF) are obtained 1.5 cm and 49.7 cm in full-wave simulations, respectively, demonstrating good agreement with the theoretical predictions. Moreover, the scanning performance of the CEDG structure is investigated by utilizing the tilting of the SM and feed point displacement, presenting a maximum field-of-view (FoV) of 40 cm at a distance of 3 m, which is acceptable for modern practical stand-off imaging systems. Finally, the capability of the proposed structure for tuning the stand-off distance by lateral displacement of the feeding source is verified. As a result, the proposed compact, low-cost configuration shows potential for practical mm-wave and THz stand-off imaging systems.

\end{abstract}

\section{Introduction}
Millimeter-wave (mm-wave) and terahertz (THz) imaging systems have found widespread application in recent years \cite{lai2024single,xu2024contraband,ismagilov2023assessment,ghamsari2022design,ning2023binary}. In particular, significant improvements have been made for mm-wave and THz stand-off imaging systems in terms of resolution, field-of-view (FoV), frame rate, stand-off distance, dimensions, and overall weight \cite{shojaeian2021improving,gatti2022standoff,prasad2021high,strahl2021methane}.

Traditionally, refractive lenses have been the main component of imaging system optical structures. In contrast, reflective mirrors can potentially eliminate various frequency-dependent losses, dispersions, and aberrations, enabling enhanced imaging performance and higher power handling that is critical for high-power laser systems \cite{korsch2012reflective,burkov2024technique,dev2022generating,christensen2021measuring,liu2024efficient,rao2023generation,stromberg2024practical}. In addition, this characteristic of reflective-based optical structures allows the implementation of wideband radar imaging techniques, such as frequency-modulated continuous-wave (FMCW), with high precision \cite{yurduseven2018point,hu2022high,grajal20153}.

Regarding technical design aspects, practical lens-based imaging systems require two different structures for the transmitter and receiver sections, noticeably increasing the overall dimension and weight of the system. In contrast, reflective-based configurations utilize a single transmitting/receiving optics \cite{tzydynzhapov2020new,zhou2018bifocal}. In addition, the use of focal-plane-arrays (FPAs), which is still a technological bottleneck in mm-wave and THz frequencies, to perform object plane scanning seems to be the only practical option in refractive-based imaging systems \cite{dhillon20172017}. On the other hand, the relatively low weight of the mirrors allows in practice a pixel-by-pixel scanning of the object by tilting a single small mirror, avoiding reliance on the FPAs. Consequently, the reduced weight of the components allows the use of larger apertures for improved resolution \cite{lee2020high,stutzman2012antenna}. Furthermore, a beam splitter can be used in reflective structures to isolate the transmitted and received beams efficiently, allowing the flexible positioning of the source and detector based on electrical and mechanical considerations \cite{rao2019reconfigurable}. Finally, the possibility of utilizing normal object illumination for obtaining higher image quality is another advantage of adopting reflective optics in imaging structures \cite{cooper2011thz}. 

Besides the mentioned pros for optical reflective structures, imaging structures consisting of mirrors suffer from several challenges both in design and performance \cite{tachi2021chromatic}. First, reflective optical configurations often require precise alignment and positioning of mirrors, complicating the design process. Second, reflective optics suffer from surface scattering and diffraction, particularly at shorter wavelengths where refractive lenses excel. Third, the performance of reflective optics is highly dependent on the quality of the mirror surfaces. Indeed, any imperfections or contaminants can significantly degrade image quality. Fourth, although reflective optics can reduce weight in smaller configurations, larger systems may require substantial structural support for mirrors. Fifth, reflective systems can introduce artifacts such as ghost images due to multiple reflections within the system. Last but not least, the optical blockage, where stray rays reflect toward the detector, can be a critical challenge in many practical mirror-based imaging structures \cite{moldsvor1992analysis,reales2024multi,yang2024shaping,pandiyan2021reflective}. To be more specific, the blockage phenomena in imaging systems can cause detector saturation, reducing sensitivity, signal-to-noise ratio (S/N), dynamic range, and isolation. Furthermore, considering the challenges of generating mm-wave and THz radiation \cite{berry2013significant}, antenna power scattering decreases source efficiency in active imaging systems. Moreover, while larger apertures enhance image resolution according to the Rayleigh equation, they also worsen blockage and diffraction effects, limiting resolution, scan range, and maximum achievable FoV \cite{saleh2008fundamentals,sabui2024impact,yang2024reconstruction,wu2022computed}. 

Many studies are dedicated in the literature to address these challenges with different optimization goals such as reducing optical aberrations, extending the FoV, desensitization of design parameters, and improving the overall efficiency of on-axis and off-axis structures \cite{wu2020design,gong2016full,meng2018desensitization,fuerschbach2014theory,yang2018aberration,gu2020optical,qin2021design,gong2017point,liu2018design,yang2014design,yang2015direct}. For instance, paper \cite{qin2021design} proposes a method to evaluate and reduce tilt error sensitivity in optical systems, demonstrating its effectiveness through wavefront error comparisons in an off-axis three-mirror system. A direct design method for three-dimensional freeform surfaces and imaging systems using a construction-iteration process is presented in \cite{yang2015direct} which significantly enhances image quality in an off-axis three-mirror system. Specifically, complicated free-form design algorithms including shaping and deforming mirrors using iterative approaches and the integration of freeform mirrors are utilized extensively to obtain more compact off-axis optical systems, reducing the optical obstruction and optimizing imaging performance while significantly reducing system volume \cite{liu2024compact,eichenberger2018deformable,huang2017wavefront,brossard2018terahertz}. For example, a design concept that integrates freeform mirrors and detectors to create more compact off-axis optical systems is introduced in \cite{liu2024compact}, demonstrating volume reduction while maintaining imaging quality. In another work, \cite{huang2017wavefront} presents a low-cost deformable mirror group for wavefront correction in small-aperture fiber lasers, consisting of nine single-actuator mirrors arranged in a 3×3 array, which offers high lateral resolution and effective aberration correction. However, these approaches suffer from several challenges, including the removal of surface symmetry, increased fabrication costs, and degradation of the imaging performance due to unequal ray spacing and different optical path lengths (OPLs). 

Applying new optical techniques to classical reflective structures can lead to improved performance for new applications. For instance, an optimized version of the traditional optical Cassegrain antenna with improved transmission efficiency is presented \cite{zhang2015hyperbola}. In \cite{llombart2010confocal,martinez2008zooming}, the optical obscuration of the on-axis reflective setups is mitigated by performing several modifications on the classical Gregorian structures. As a result, a new aberration-free off-axis setup with improved imaging performance is obtained for the optical section of mm-wave and THz imaging systems. However, the proposed configurations suffer from complex mirror alignments and large physical dimensions. 

In this paper, we propose a simple compact symmetric on-axis dual-mirror confocal ellipsoidal displaced Gregorian configuration, called CEDG hereafter, with reduced blockage characteristics for stand-off imaging systems. It is worth noting that while our proposed methodology in designing the CEDG structure is not inherently limited to Gregorian configurations and a similar design approach could be implemented in a Cassegrain system, it would require placing additional mirrors between the feed and sub-mirror, increasing design complexity and feed blockage risk. To be more specific, we selected Gregorian geometry as the baseline for our proposed method due to its advantageous optical path characteristics, particularly the placement of the secondary mirror beyond the primary focus, which facilitates the integration of a compound secondary mirror without significant obscuration or alignment issues. This configuration supports a folded optical path, enabling more compact system designs, an especially important consideration in the mm-wave and THz applications where component alignment and wavefront quality are critical. Additionally, the Gregorian layout offers better accessibility for positioning focal plane components behind the primary mirror, reducing blockage and easing system packaging. Therefore, the Gregorian system was chosen as a suitable candidate for demonstrating the concept due to the reasons mentioned above.

In Section \ref{section.2}, the general geometry of our proposed CEDG structure is introduced, followed by a step-by-step design guide for constructing the modified secondary reflective surface. Next, geometrical optics principles are applied and the required design formulas are extracted in Section \ref{section.3}. In Section \ref{section.4}, the blockage reduction characteristic of the proposed structure is investigated by comparing the CEDG configuration ray tracing results with a Gregorian structure. Finally, the imaging performance of the proposed optics is discussed in detail in Section \ref{section.5} in terms of resolution, depth-of-focus (DoF), FoV, and stand-off distance tunability. Paper results are concluded in Section \ref{section.6}.

\section{DESIGN OF THE PROPOSED CEDG STRUCTURE} \label{section.2}

\subsection{General geometry}
\begin{figure}[htbp]
\centering\includegraphics[width=15cm, trim={2.8cm 18.3cm 1cm 3.5cm},clip]{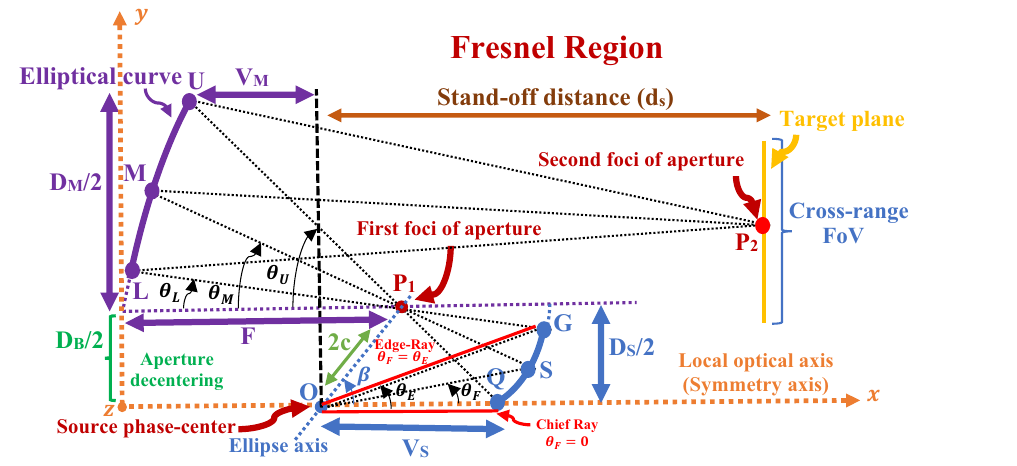}
\caption{General geometry of the CEDG setup at the $xy$ plane. Emanated divergent rays from the point $O$, after reflection from the SM, are focused on the target plane at point $P_2$ by reflection from the main ellipsoidal MM, which is decentered vertically $D_B/2$, leaving a hole in the middle of the geometry. Conversely, based on optical reciprocity, the rays that are emanated from point $P_2$ on the target plane, are focused at the point $O$.}
\label{fig.1}
\end{figure}

The geometrical structure of the proposed CEDG configuration is illustrated in Fig. \ref{fig.1}. Unless otherwise stated, all geometrical parameters and symbols defined in this section follow standard optical design conventions, particularly with respect to sign conventions for angles, distances, and coordinate directions. $D_M$ and $D_S$ denote the diameters of the main mirror (MM) and secondary mirror (SM), respectively. The MM curve is a standard elliptical conic section, while the SM is constructed of a dual semi-elliptical curve. The MM curve is vertically displaced by $D_B/2$. The first focal point of the aperture is shown by $P_1$, positioned at the common first focal point of the SM. Also, the second focal point of the aperture denoted by $P_2$, is located at the desired stand-off distance of $d_S$ in the target plane. $F$ denotes the focal length of the MM. The SM curve is generated using an ellipse with eccentricity $e$ and interfocal distance $2c$. Angles $\theta_U$ and $\theta_L$ define the upper and lower angles of the MM, respectively. The SM edge angle is represented by $\theta_E$ and $\beta$ is the tilt angle between the optical axis and the ellipse axis. We also define an auxiliary feed ray path $\theta_F$ in the $z=0$ plane, assuming $|\theta_F |\leqslant|\theta_E |$, with a corresponding MM angle $\theta_M$. Moreover, the corresponding chief ray ($\theta_F=0$) and the SM edge-ray ($\theta_F=\theta_E$) are also illustrated. The elliptical aperture can focus the rays reflected from the SM onto the target plane. By definition, this second focal plane of the MM can be positioned at the near-field (Fresnel region) of the aperture.

The source feeding point is located at point $O$, which is the on-axis second focal point of the SM. The other focal point of the SM coincides with the focal point of the aperture at point $P_1$. Parameters $V_M$ and $V_S$ denote the x-coordinates of the MM and SM points, respectively, corresponding to the feed principal ray of the optical geometry. $V_S$ and $V_M$ are two geometrically important parameters as they indicate the relative position of the physical source with respect to the positions of the MM and SM. Notably, $V_S$ is always positive, but $V_M$ can be negative depending on the position of the MM focal point. It is worth noting that in the proposed configuration, the lowest point of the MM and the highest point of the SM are aligned along the parabolic axis. This represents a special case chosen deliberately to minimize blockage by reducing the radiation from the MM toward the SM surface. In general, the alignment of these points may vary depending on the specific design parameters and system constraints. Finally, the 3D mirror surfaces are obtained by spinning the generating curves about the x-axis (symmetry axis).

\subsection{Step-by-step procedure of constructing the SM surface}
Here, a unified construction procedure for the SM curve is proposed. Intuitively, the key point behind the proposed geometry involves applying appropriate decentering and tilting techniques to the constructive curves of a conventional offset Gregorian-based geometry to achieve the final blockage-free optical configuration. First, consider a standard elliptical curve, defined as a closed conic section obtained by tracing a linear closed path along the intersection of an oblique plane with a cone, as depicted in step (1) of Fig. \ref{fig.2}. The detailed construction procedure is as follows:

\begin{figure}[htbp]
\centering\includegraphics[width=16cm, trim={8cm 17cm 1cm 3cm},clip]{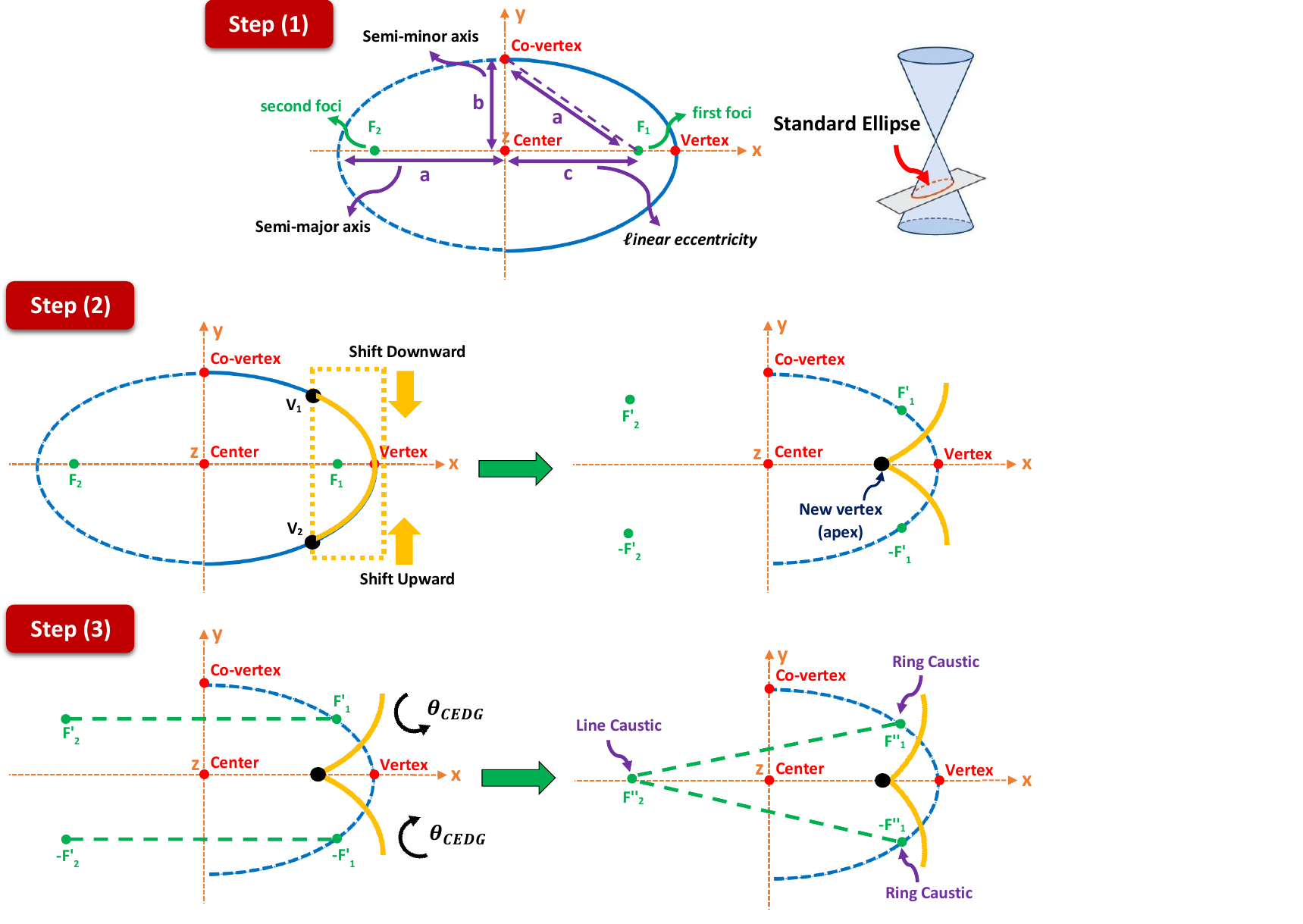}
\caption{Step-by-step procedure of constructing the SM curve.}
\label{fig.2}
\end{figure}

1- Consider a standard horizontal elliptical conic section with two on-axis foci, $F_1$ and $F_2$. The first foci $F_1$, also called the virtual foci, is coincident with the first focal point of the elliptical MM aperture, which is denoted by point $P_1$ in Fig. \ref{fig.1}. Parameters $a$ and $b$ represent the semi-major and semi-minor axis lengths of the ellipse, respectively. Parameter $c$ is defined as the distance between each focus and the central point of the curve, also called the focal distance or linear eccentricity of the ellipse, which always satisfies $c=\sqrt{a^2-b^2}$. 

2- Select one quadrant of an ellipse, equal to a fraction of the semi-major and semi-minor axis lengths in the $x$ and $y$ directions, respectively, where $V_1$ and $V_2$ are the $y$-coordinates of the intersection points of the selected section with the ellipse circumference. Shift the upper (lower) section downward (upward) so that points $V_1$ and $V_2$ are located at the same coordinates on the $x$-axis (symmetry axis). This step creates two distinct ellipses.

\begin{figure}[htbp]
\centering\includegraphics[width=13.5cm, trim={7cm 12.5cm 8cm 3.5cm},clip]{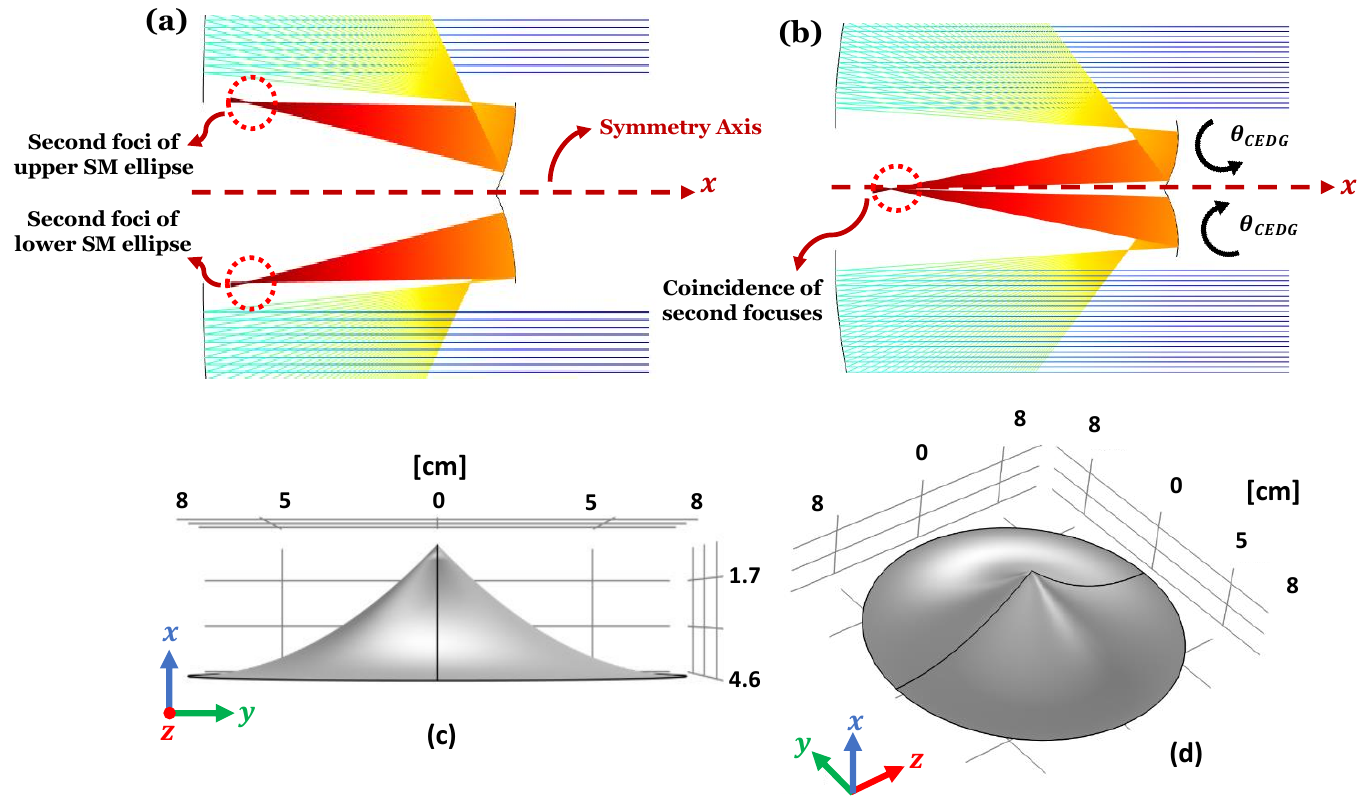}
\caption{(a) Before applying the final rotation, the second foci of the upper and lower SM ellipses are at different locations. (b) After applying the final rotation, the second foci of both ellipses converge to a single common point on the symmetry axis. (c) and (d) show the two- and three-dimensional cross-sections of the constructed SM surface, respectively.}
\label{fig.3}
\end{figure}

3- Rotate the upper (lower) ellipse counter-clockwise (clockwise) around the $z$-axis, which is orthogonal to the $xy$-plane, such that foci $F'_1$ of the upper and lower ellipse are placed at the first focal point of the MM curve (point $P_1$ in Fig. \ref{fig.1}) and the second foci $F'_2$ of both ellipses coincident at a single common point on the symmetry axis, denoted by point $O$ in Fig. \ref{fig.1} and is the feeding source point position of the structure. Notice that in practice it is easier to apply the rotation with respect to point $F'_1$, therefore $F'_1=F''_1$. As can be more clearly observed in Fig. \ref{fig.1}, the required rotation angle is given by:

\begin{equation}
    \label{eq.1}
    \theta_{CEDG} = \arcsin(\frac{D_B}{4c}).
\end{equation}

\noindent where $2c$ is the interfocal length of the elliptical MM curve and $D_B$ is the aperture decentering length defined in Fig. \ref{fig.1}. This rotation must also be applied to the MM curve. Also, to further illustrate the purpose of the final rotation, one can observe Fig. \ref{fig.3}(a) and \ref{fig.3}(b) which show the ray diagram of the SM section of the CEDG structure before and after rotating the upper and lower elliptical curves of the SM. The MM curve is also obtained by defining a displaced aperture of an elliptical conic section where its first focal point is located at the position of the first foci of the corresponding upper or lower SM ellipse. Finally, regarding the rotational symmetry of the geometry, the 3D surfaces of revolution are obtained by spinning the generated curves about the symmetry axis. For the MM curve, the generated one-half of the ellipse is rotated not around the symmetry axis at the vertex, but rather with the vertex traveling in a ring around a cylinder with the same diameter as the SM. Therefore, the SM first foci form a ring caustic, and the SM second foci form a linear caustic. To be more specific, the junction between the two elliptical segments of the SM introduces a non-smooth point (or cusp) in the surface profile. This cusp generates a ring-shaped beam that predominantly appears in the near-field region, and its intensity particularly depends on the optimized design conditions. Although the cusp introduces localized phase and amplitude variations, these effects can be effectively mitigated through careful alignment and appropriate selection of the tilt angle between the mirror segments. In some configurations, the field shaping induced by the cusp may be advantageous, enabling the suppression of side lobes or improved power distribution across the aperture. Theoretically, the larger tilt angles tend to enhance the discontinuity and its associated optical effects, while smaller angles promote smoother field transitions and diminish the cusp’s influence.

The final side-view and perspective-view of the constructed SM surface are shown in Fig. \ref{fig.3}(c) and \ref{fig.3}(d), respectively. It is also worth noting that, from the physical optics point of view, the obtained SM surface can theoretically convert electromagnetic waves with spherical wavefronts from the point $O$ (see Fig. \ref{fig.1}) to a uniform phase field distribution on the aperture.

\section{GEOMETRICAL-BASED ANALYTICAL STUDY} \label{section.3}
In order to extract closed-form design equations, an analytical study is performed. A more detailed trigonometric analysis of the SM structure is provided in the supplementary section. Based on Fig. \ref{fig.1}, an ellipse generating the MM curve can be described as:

\begin{equation}
	\overline{P_1 M} = \frac{2a}{1 + \cos\theta_M}, \label{eq2}\\
\end{equation}

Moreover, considering an elliptical conic section of the MM, using the standard definition of an ellipse:

\begin{equation}
	\overline{P_2 M} \mp \overline{M P_1} = 2a, \label{eq3}\\
\end{equation}

This equation describes the distance from one focus $P_2$ to a point $M$ on the ellipse and then to the other focus $P_1$. Thus, an elliptical MM can be described in its most general form as:

\begin{equation}
	\overline{P_2 M} \mp \overline{M P_1} = \frac{2c}{e}.  \label{eq4}\\
\end{equation}

Similarly, the SM curve can be described as:

\begin{equation}
	\overline{O S} \mp \overline{S P_1} = \frac{2c}{e}. \label{eq5}\\
\end{equation}

Also, one can directly observe from Fig. \ref{fig.1}:

\begin{equation}
	\tan{\theta_U} = \frac{-D_M}{2(V_S - V_M)}. \label{eq6}\\
\end{equation}

Following a similar mathematical approach in the literature \cite{moreira2001generalized}, from Fig. \ref{fig.1}, the constant OPLs of the chief ray ($\theta_F=0$) and SM edge-ray ($\theta_F=\theta_E$) can be expressed as: 

\begin{equation}
	l_0 = (V_S-V_M) (1+\frac{1}{\cos{\theta_U}}), 
    \label{eq_l0_1}\\
\end{equation}

\begin{equation}
	l_0 = \frac{D_S}{2\sin{\theta_E}}-\frac{D_S}{2\tan{\theta_E}}-\frac{D_B-D_S}{2\sin{\theta_L}}-\frac{D_B-D_S}{2\tan{\theta_L}}, 
    \label{eq_l0_2}\\
\end{equation}

respectively. Then, angle $\theta_U$ can be calculated by combining Eqs. (\ref{eq6}) and (\ref{eq_l0_1}):

\begin{equation}
	\tan{(\theta_U/2)} = \frac{-D_M}{2l_0}, \label{eq8}\\
\end{equation}

Also, from Eq. (\ref{eq_l0_2}), the angle $\theta_L$ can be calculated as:

\begin{equation}
	\tan{(\theta_L/2)} = \frac{-D_B+D_S}{2l_0-D_S\tan{(\theta_E/2)}}, \label{eq7}\\
\end{equation}

\noindent where the distance between the SM and MM surfaces is assumed to be approximately $l_0/2$. Next, applying the law of sines to triangle $OP_1Q$ in Fig. \ref{fig.1} yields:

\begin{equation}
	\frac{\sin(\theta_U)+\sin(\theta_L)}{\sin(\beta)+\sin(\beta-\theta_U)} = \frac{-\sin(\theta_E-\theta_L)}{\sin(\beta-\theta_L)+\sin(\beta+\theta_E)}, \label{eq9}\\
\end{equation}

\begin{equation}
	V_S = \frac{(-D_S/2)\sin(\theta_E-\theta_L)\sin(\beta-\theta_U)}{\sin(\theta_E)\sin(\theta_U)\sin(\beta-\theta_L)}. \label{eq10}\\
\end{equation}

\noindent which can be solved simultaneously to find $\beta$ and $V_S$. With the values of $\theta_L$, $\theta_U$, $V_S$ and $\beta$, the interfocal distance $2c$ and the eccentricity $e$ of the SM can be obtained as:

\begin{equation}
	2c = \frac{-V_S\sin(\theta_U)}{\sin(\beta-\theta_U)}, \label{eq11}\\
\end{equation}

\begin{equation}
	e = \frac{-\sin(\theta_U)}{\sin(\beta)+\sin(\beta-\theta_U)}. \label{eq12}\\
\end{equation}

\noindent which uniquely determines the SM surface. Moreover, the following constraints on the pre-defined constants and angles can be observed:

\begin{equation}
\left\{
\begin{array}{l}
0 < e < 1, \\
0 < \beta < \pi, \\
0 < \theta_E < \pi, \\
0 \leq \theta_F \leq \theta_E, \\
\theta_L < 0.
\end{array}
\right.
\label{eq13}\\
\end{equation}

By adopting the geometrical optics (GO) principles, three possible blockage mechanisms of the proposed CEDG structure can be analyzed. The SM blockage, specified by the incidence of the MM reflected rays upon the SM, is avoided when:

\begin{equation}
    D_S\leq D_B<D_M.
\label{eq14}\\
\end{equation}

Furthermore, the feed blockage, which happens when parts of the MM reflected rays incident upon the feed structure, never occurs in the proposed CEDG configuration, as a point source is considered for the feeding. Similarly, self-blockage, which occurs when the SM lower (upper) half reflected rays intersect with the SM upper (lower) half surface, never happens in the proposed CEDG geometry.

\section{DESIGN EXAMPLE} \label{section.4}
In this section, a design example is presented to illustrate the design procedure and to examine the characteristics of the CEDG structure. To be more specific, the following steps are recommended for the design of the proposed CEDG optics:

\noindent 1- Define the design parameters based on the desired specifications, physical dimensions, and constraints.

\noindent2- Calculate the geometrical and optical parameters from the equations of Section \ref{section.3}.

\noindent3- Compare the calculated parameters with the physical considerations and blockage constraints of Eqs. (\ref{eq13}) and (\ref{eq14}). Repeat the calculation if they violate the constraints.

\noindent4- Use Eqs. (\ref{eq2})-(\ref{eq5}) and follow the construction procedure in Section \ref{section.2} for designing the MM and SM surfaces of the CEDG configuration based on the parameters calculated in the previous steps. One can also use the trigonometric definitions and guidelines to better design the structure by defining the coordinates of different points on the geometry (see Supplementary section).

Following this design procedure, first, the parameters of Table \ref{Design Param} are selected for constructing the CEDG optical setup.

{\rowcolors{1}{cyan!10}{}
\begin{table}[htbp]
	\centering
	\caption{\bf Design parameters of the CEDG structure.}
	\begin{tabular}{|c|c|c|c|c|c|c|c|}
		\hline
		\textbf{$D_M$} & \textbf{$D_S$} & \textbf{$D_B$} & \textbf{$\theta_E$} & \textbf{$l_0$} & $d_s$ & FoV & Resolution \\ 
		\hline
		80 cm    & 16 cm           & 16 cm           & $15^{\circ}$            & 35 cm &      3 m     & 40 cm &       1.5 cm           \\ 
		\hline
	\end{tabular}
	\label{Design Param}
\end{table}
}

Parameters $l_0$ and $\theta_E$ are chosen to have a blockage-free geometry based on the discussions in section \ref{section.3}. The parameter $D_B$ is also selected to be equal to $D_s$ for maximum aperture efficiency. In order to minimize the blocking effects of the SM, its diameter can be chosen much smaller than the aperture diameter. The MM is defined as an elliptical curve with its second foci focused at the stand-off distance $d_s$=3 m, which is a conventional imaging distance in commercial stand-off imaging systems. The calculated parameters of the MM and SM conic sections, together with some other useful optical constants are summarized in Table \ref{optical param}.

\begin{table}[htbp]
	\centering
	\caption{\bf Calculated geometrical and optical parameters of the CEDG structure design example.}
	\begin{tabular}{|c|c|c|c|c|c|c|c|}
		\hline
            \rowcolor{cyan!10}
		 & \textbf{$F_1,F_2$} & \textbf{$a$} & \textbf{$b$} & \textbf{$c$} & \textbf{$e$} & $M_e$ & $\kappa$ \\ 
		\hline
             \cellcolor{cyan!10}
		\textbf{MM curve} & 300, 3000 mm    & 1650 mm           & 948.6 mm           & 1350 mm            & 0.8 &      9.5     & -0.65        \\ 
		\hline
            \cellcolor{cyan!10}
		\textbf{SM curve} & 50, 300 mm    & 175 mm           & 122.4 mm           & 125 mm            & 0.7 &      5.6     & -0.4        \\ 
		\hline
	\end{tabular}
	\label{optical param}
\end{table}

$F_1$ and $F_2$ are the first and second focal lengths of the MM and SM, respectively. Parameters $a$, $b$, and $c$ are the semi-major axis, the semi-minor axis, and half of the focal length (linear eccentricity) of the ellipses, respectively. Parameter $e=c/a$ is the eccentricity, which is unity for a parabolic curve and less than unity for an elliptical curve. Also, parameter $M_e$ in this table is defined as the magnification of the ellipse and can be calculated as:

\begin{equation}
    \label{eq.15}
    M_e=\frac{1+e}{1-e},
\end{equation}

The constant $\kappa$ is defined as:

\begin{equation}
    \label{eq.16}
    \kappa=p-1=-e^2,
\end{equation}

\noindent where $p$ is the conic constant of the ellipse. It is also worth noting that these definitions are based on the standard elliptical curve, which satisfies the following conic section equation in the yz-plane passing through the origin:

\begin{equation}
    \label{eq.17}
    z=\frac{cy^2}{1+\sqrt{1-pc^2y^2}}.
\end{equation}

\begin{figure}[htbp]
\centering\includegraphics[width=13cm, trim={4cm 6cm 3.3cm 4cm},clip]{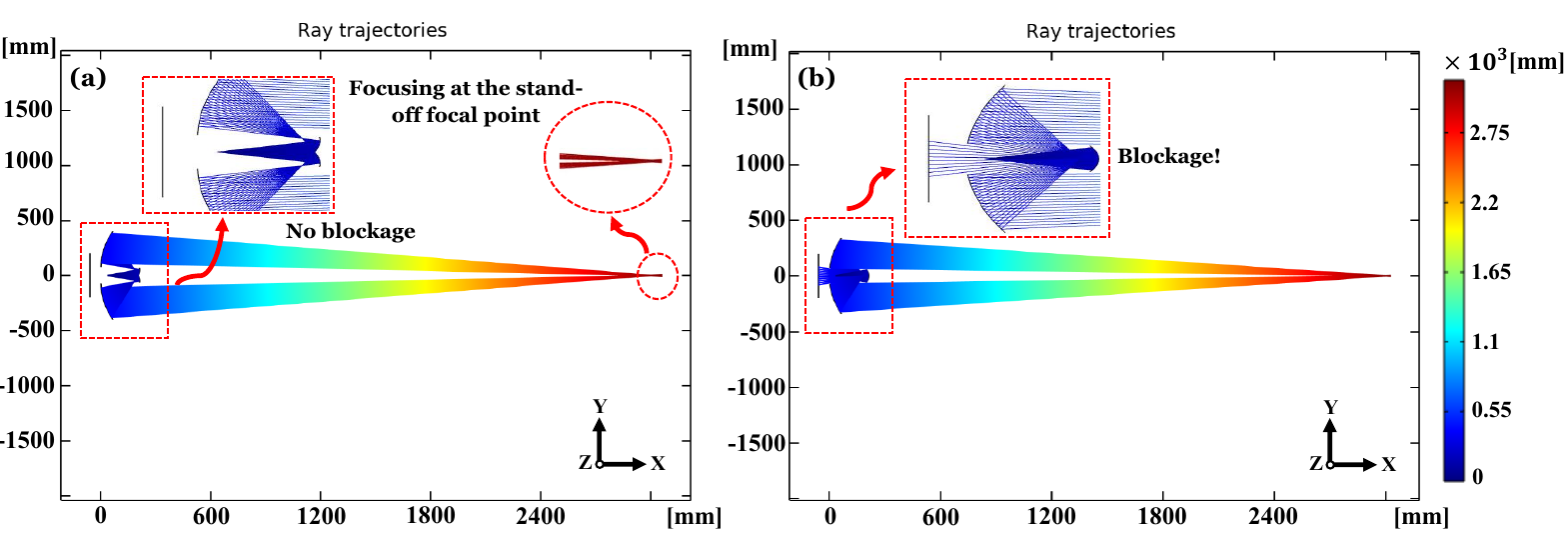}
\caption{Comparison of ray tracing results of (a) the proposed CEDG configuration and (b) the classical Gregorian structure. Both configurations are in the focusing transmitting optics mode and the rays emanating from the source point are focused at a distance of $3$ m. The color bar shows the OPL of the rays, calculated from the feeding focal point of the structure from which the rays are emitted. In comparison to the proposed CEDG design, the ray tracing results of the conventional Gregorian-based configuration show significant blockage in the source or detector region.}
\label{fig.4}
\end{figure}

It is also useful to note that manufacturing constraints and imaging performance trade-offs can also be considered in the selection of these parameters. The CEDG geometry is constructed using the calculated parameters. The Ray Optics module of the COMSOL Multiphysics is utilized to perform the ray tracing algorithm and the results are illustrated in Fig. \ref{fig.4}. In this figure, the rays have been colored according to the OPL of the rays calculated from the source point. It can be seen that no rays pass through the central blockage region of the proposed CEDG structure. Therefore, the blockage is eliminated in the CEDG structure. Furthermore, the reflected rays from the MM are focused at the second out-of-structure focal point which can be considered to be the stand-off distance of the imaging system.

\section{IMAGING PERFORMANCE OF CEDG STRUCTURE} \label{section.5}

\subsection{Resolution}
\begin{figure}[htbp]
\centering\includegraphics[width=13.5cm, trim={11cm 13.5cm 10cm 0.5cm},clip]{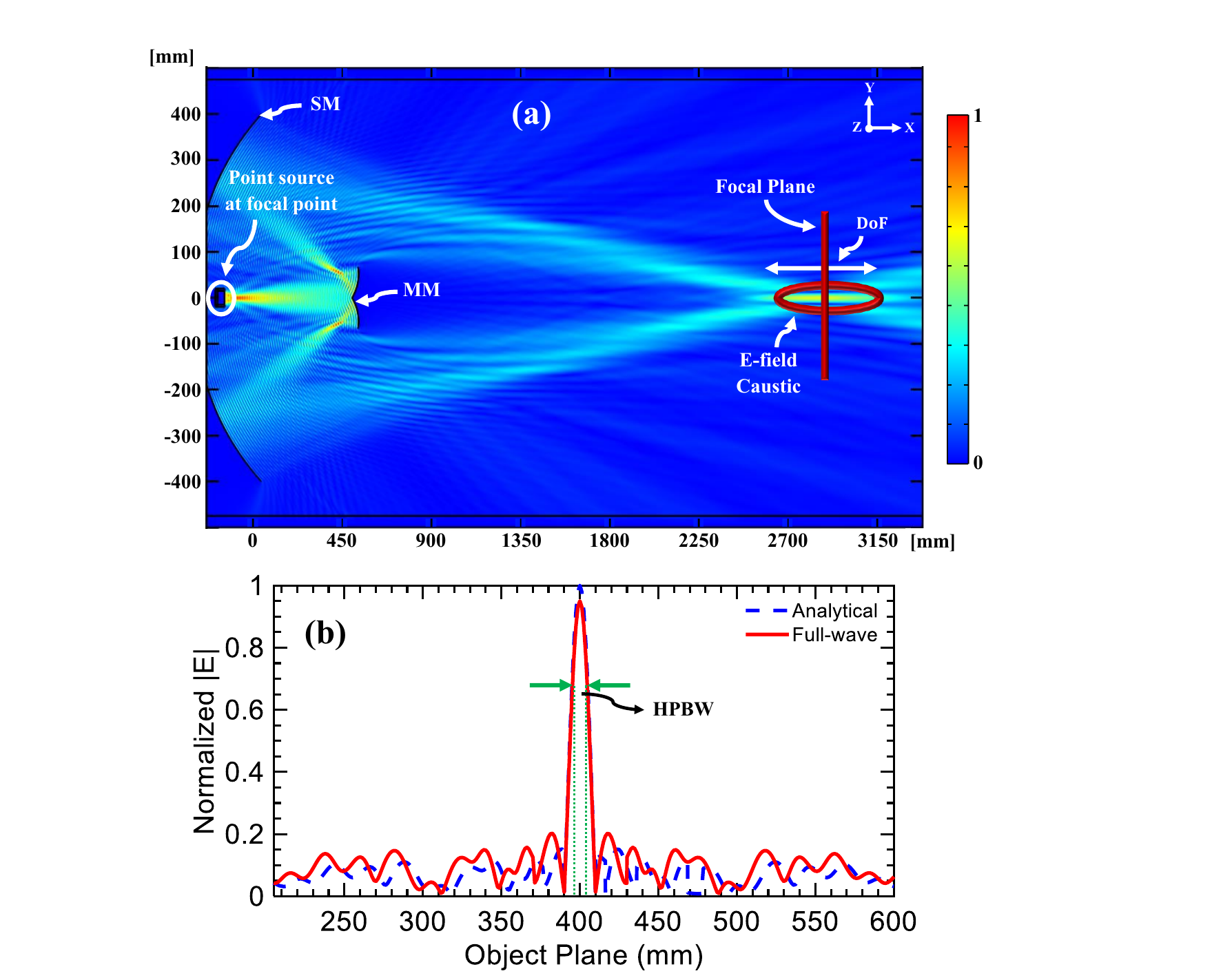}
\caption{(a) Full-wave simulation results of the E-field intensity distribution at the Fresnel region of the designed CEDG structure in the xy-plane. A point source excitation is used at the feeding focal point at the central frequency of $60$ GHz. The stand-off focal plane and the obtainable DoF as a result of the created E-field caustic are also illustrated. (b) Calculated normalized E-field intensity distribution at the stand-off focal plane located at a distance of 3 m from the mirror setup using both the full-wave simulation and Eq. (\ref{eq.18}).}
\label{fig.5}
\end{figure}

To investigate the realized resolution of the designed CEDG structure of Section \ref{section.4}, the point spread function (PSF) can be calculated. First, the electric field (E-field) distribution at the CEDG structure Fresnel region is obtained by performing full-wave simulations using the Electromagnetic Waves module of the COMSOL Multiphysics, and the result is shown in Fig. \ref{fig.5}(a). Also, suitable absorbing boundary conditions are applied near the simulation medium to avoid unwanted reflections. 

By definition of the PSF, a point source at the central frequency of $60$ GHz is defined at the feeding focal point of the structure. For imaging in the Fresnel region, the object plane should be located at the secondary focus of the elliptical MM. In this case, an analytical expression of the electric field on the object plane can be expressed as a series of incident plane waves. Following this approach, the $z$-component of the electric field on the focal plane can be described in the general form as \cite{rao2013handbook}:

\begin{equation}
    \label{eq.18}
     E_f^z(y,z)=\frac{-je^{-jkd_s}e^{-jk\frac{y^2+z^2}{2P_1}}}{2\pi d_s}\iint_{S_r}{e(y',z')m_y(y',z')k_xe^{jk\frac{y'y}{d_s}}e^{jk\frac{z'z}{d_s}} \, dy'dz'}.
\end{equation}

\noindent where $d_s$ is the stand-off distance at which the object plane is located, $P_1$ is the first focus of the MM, $k=2\pi/\lambda$ is the wave number,
$S_r$ is the integration domain over the aperture surface, $e(y',z')$ is the equivalent aperture field generated by a series of plane waves coming from the first focus, and $m_y (y',z')$ denotes the polarization effect on the aperture field. The calculated 1-D normalized E-field intensity at the stand-off focal plane of Fig. \ref{fig.5}(a) using both the full-wave simulations and Eq. (\ref{eq.18}) is illustrated in Fig. \ref{fig.5}(b). The peak of the E-field intensity distribution in both results is located at a distance near 3 m from the structure, as expected by comparing with the ray tracing results of Fig. \ref{fig.4}(a), and there is a good agreement between the results obtained from both methods.

The half-power beamwidth (HPBW) of the PSF pattern can be considered as a measure of the smallest achievable resolution that can be obtained with an active imaging system when the transmitter and receiver patterns are the same, which is usually referred to as a diffraction-limited imaging system. Based on Fig. \ref{fig.5}(b), the HPBW value is calculated to be approximately $15$ mm. Furthermore, considering the point source excitation frequency of 60 GHz, corresponding to a wavelength of $5$ mm, the achievable resolution using the designed reflective imaging structure can be approximated as \cite{rao2013handbook}:

\begin{equation}
    \label{eq.19}
    Resolution=0.74 \frac{\lambda \times d_s}{D_M}
.
\end{equation}

which is calculated approximately $13.8$ mm for our designed CEDG configuration. By taking into account the effect of the pattern sidelobes, a relatively good agreement between the theoretical calculation and the simulation results can be observed.

\subsection{Depth-of-focus (DoF)}
When utilizing an elliptical mirror as the main aperture of an imaging structure, one may observe that the stand-off focal point exhibits a localized maximum along the axial direction prior to being well-focused. This observation is referred to as focal shift, and the area where the field remains properly focused is defined as the DoF of the imaging system \cite{li1981focal}. Through the use of full-wave simulations, the DoF for a reflective imaging system can be defined as the axial region in which the field decay is less than 3 dB compared to the maximum at exactly the stand-off distance. In practical terms, this value represents the region in which an object can be placed without experiencing a significant reduction in imaging quality. The DoF value of the designed reflective CEDG structure can be approximated as \cite{rao2013handbook}:

\begin{equation}
    \label{eq.20}
    DoF=1.77 \frac{d_s}{N}.
\end{equation}

\noindent where $N=D_M^2/4\lambda d_s$ is the Fresnel number of the system. For our design, the DoF is calculated $49.7$ cm based on this equation. The theoretically predicted value of DoF has an excellent agreement with the obtained DoF value from Fig. \ref{fig.5}(a) which is about $49.3$ cm.

\subsection{Scanning performance and FoV}
In this section, two methods have been utilized to evaluate the stand-off focal plane scanning ability of the designed CEDG structure of Section \ref{section.4}.

\noindent 1- \textbf{Tilting of the SM}: Using the tilt angle of a small rotating mirror is a popular and relatively simple technique for scanning the object plane in THz and mm-wave imaging systems. Moreover, this can potentially eliminate the requirement of using costly FPA detectors. 

To have an intuition on the required angular displacement of the SM surface for scanning the object plane, one can calculate the beam deviation factor (BDF), defined as the ratio of the angular deviation of the beam from its original trajectory to the initial angle of incidence of the beam. The BDF helps to understand how tilting the SM affects the focused beam's direction and quality. Specifically, by calculating the BDF when tilting the SM, it is possible to determine how much the focused beam shifts at a given stand-off distance (e.g., 3 m). This helps in predicting how effectively the imaging structure can scan across a desired area. A low BDF indicates that the beam remains close to its intended path, which is essential for maintaining high gain and efficiency. 

The BDF of a reflective scanning structure varies depending on the applied offset and f-number of the optical system. First, the deviation angle of rays with respect to the optical axis can be calculated as:

\begin{equation}
    \label{eq.21}
    \theta_{sb}=\arctan \frac{FoV}{2\times d_s},
\end{equation}

\noindent which is calculated roughly $3.81^{\circ}$ for our design. Then, the BDF value for the maximum SM tilting of $\theta_s=4.5^{\circ}$ can be determined as:

\begin{equation}
    \label{eq.23}
    BDF=\frac{2\times \theta_s}{M_s \theta_{sb}}.
\end{equation}

\noindent where $\theta_s$ is the tilting angle of the SM and $M_s$ is the magnification of the optical setup and is defined as the ratio of the first focal length of the MM to the first focal length of the SM. Based on Eq. (\ref{eq.23}), the BDF of the designed CEDG structure of Section \ref{section.4} is roughly equal to 0.39. Technically, a BDF between 0.1 and 0.5 is acceptable in scanning imaging systems where slight distortions do not significantly impact overall performance.

\begin{figure}[htbp]
\centering\includegraphics[width=15cm, trim={17cm 10cm 10cm 0.5cm},clip]{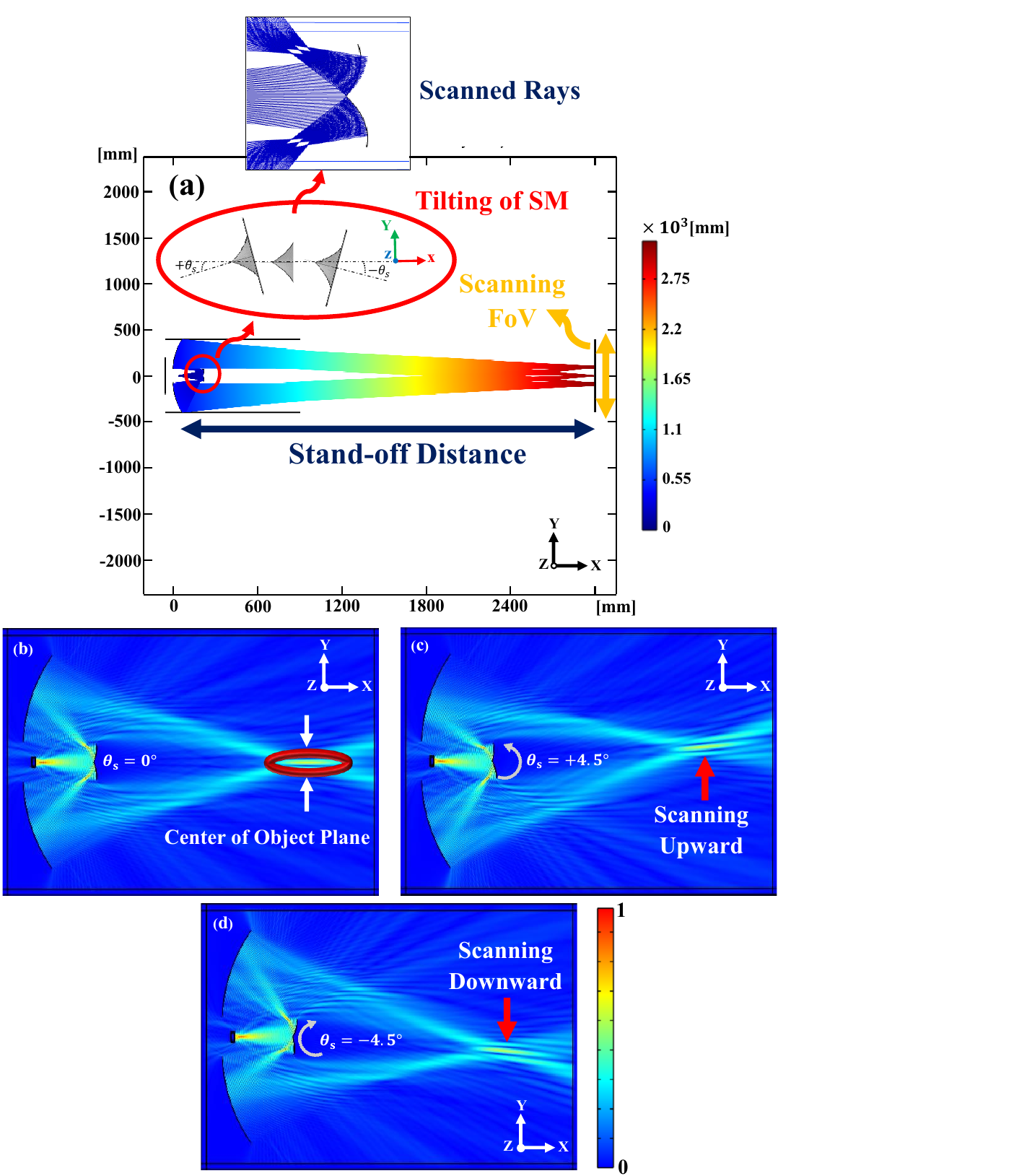}
\caption{Performing stand-off focal plane scanning by tilting the SM. (a) Ray tracing simulation results of the designed CEDG structure in the xy-plane, illustrating the scanning of rays on the stand-off object (focal) plane by the axial tilting of the SM. The full-wave simulation results of the normalized E-field intensity at the Fresnel region of the designed CEDG structure in the xy-plane, showing the stand-off object plane scanning performance for (b) $\theta_s=0^{\circ}$, (c) $\theta_s=+4.5^{\circ}$, and (d) $\theta_s=-4.5^{\circ}$.}
\label{fig.6}
\end{figure}

\begin{figure}[htbp]
\centering\includegraphics[width=14cm, trim={2.2cm 7.5cm 1cm 1cm},clip]{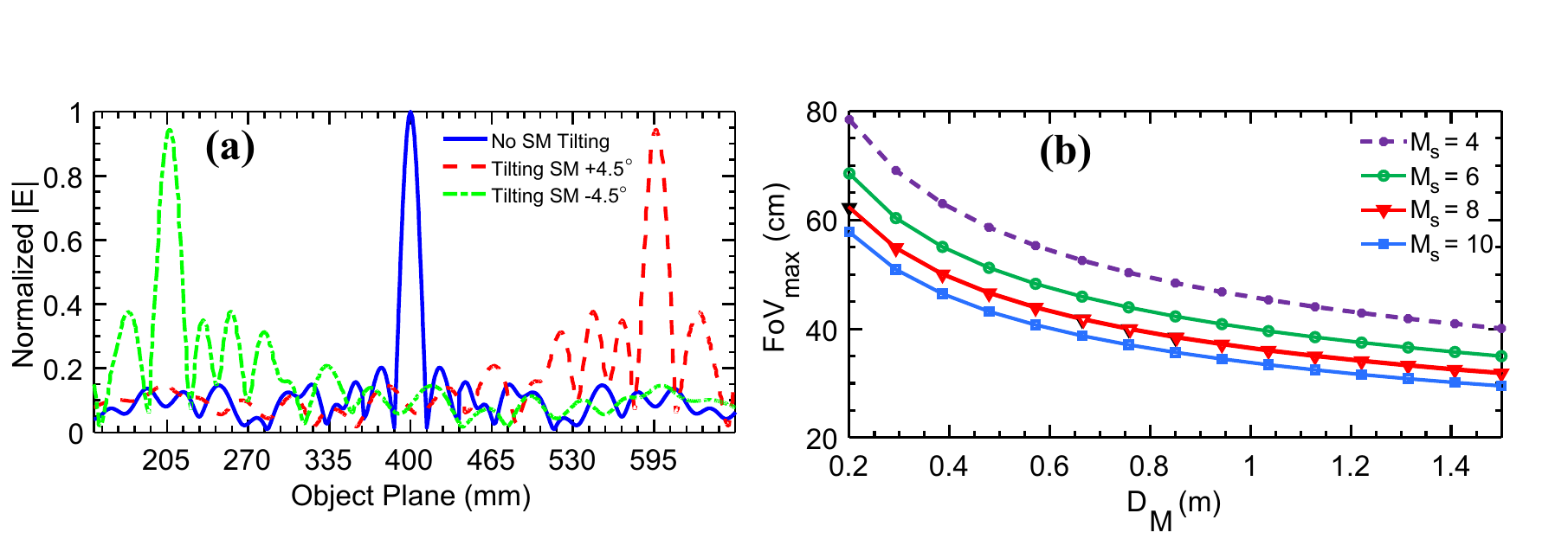}
\caption{(a) Calculated normalized E-field intensity distribution on the stand-off focal plane of the CEDG structure for the three SM tilting states of $\theta_s=0^{\circ}$, $\theta_s=+4.5^{\circ}$, and $\theta_s=-4.5^{\circ}$. The HPBW of the main lobe of the focused beam is increased as a result of diverging the emanated rays during the scanning. (b) Changes of maximum achievable FoV in the designed CEDG configuration of Section \ref{section.4}, calculated based on Eq. (\ref{eq.24}), for $d_s$= 3 m and several magnification values of $M_s$= 4, 6, 8, 10.}
\label{fig.7}
\end{figure}

The utilization of this technique for scanning the stand-off object plane is illustrated in Fig. \ref{fig.6}. In the ray tracing results of Fig. \ref{fig.6}(a), it is shown that by tilting the SM equal to $\theta_s$ around the $z$-axis, the emanated rays are still focused on the object plane but the focusing point is shifted axially on the object plane. In addition, the full-wave simulation results of tilting the SM for the three values of $\theta_s=0^{\circ}$, $\theta_s=+4.5^{\circ}$ and $\theta_s=-4.5^{\circ}$ are depicted in Figs. \ref{fig.6}(b)-(d). First, Fig. \ref{fig.6}(b) shows the E-field distribution at the Fresnel region when the SM is not tilted and the E-field is maximized at the center of the object plane and along the optical (symmetry) axis. When the SM is tilted counterclockwise and clockwise, as depicted in Fig. \ref{fig.6}(c) and \ref{fig.6}(d) respectively, the E-field caustic is scanned vertically upward and downward, respectively. Furthermore. the normalized E-field intensity for the three states of scanning the object plane is calculated similarly to the previous sections and is demonstrated in Fig. \ref{fig.7}(a). It can be observed that due to changing the optical path of the rays by the scanning, some of the rays have diverged from their principal optical path and, as a result, the E-field pattern sidelobe levels (SLL) and also the HPBW of the main beam of rays is increased on the object plane when the rays are scanned by tilting the SM.

The tilting angle of the SM and the corresponding beam scanning area on the target plane is tabulated in Table \ref{Tilting}. Based on this table, by tilting the SM about $\theta_s=\pm4.5^{\circ}$, the focal plane on the stand-off distance of 3 m is scanned from $11.7\backsim38.93$ cm which gives approximately the desired FoV of Table \ref{Design Param}.

\begin{table}[htbp]
	\centering
	\caption{\bf Scanning the object plane in the CEDG structure using the SM tilting.}
	\begin{tabular}{|c|c|c|c|c|c|c|c|c|c|}
		\hline
              \rowcolor{cyan!10}
		\textbf{Tilt angle $\theta_s [^{\circ}]$} & $\pm0.5$ & $\pm1$ & $\pm1.5$ & $\pm2$ & $\pm2.5$ & $\pm3$ & $\pm3.5$ & $\pm4$ & $\pm4.5$ \\ 
		\hline
            \cellcolor{cyan!10}
		\textbf{Scanning range [cm]} & 11.7 & 15.4 & 19.1 & 22.6 & 26.1 & 29.6 &  32.9 & 36.1 & 38.9 \\ 
		\hline
	\end{tabular}
	\label{Tilting}
\end{table}

In mm-wave and THz imaging system utilizing lenses or mirrors, the diameter of the main lens or mirror (aperture) plays a crucial role in determining the FoV. Generally, increasing the aperture size leads to a reduction in the system's FoV. This relationship arises because the FoV is primarily governed by the focal length and sensor size. For a fixed sensor size, the FoV can be approximated by the ratio of the sensor size to the focal length. When the aperture size increases, the focal length must also increase to maintain a diffraction-limited system, which consequently reduces the FoV. Furthermore, a larger aperture significantly enhances the system's resolution by reducing the beamwidth and improving angular resolution. However, this improvement comes at the cost of a narrower FoV, as the increased aperture focuses incoming rays into a smaller angular region. Conversely, a smaller aperture results in a wider FoV but degrades resolution. This trade-off between FoV and resolution is a key design consideration in mm-wave and THz imaging systems, as achieving high-resolution imaging often necessitates sacrificing a wide field of view.

Following the approach reported in the literature for the ellipsoidal focusing structures \cite{rao2013handbook, garcia2012bifocal}, and considering a uniform aperture field distribution, the maximum achievable FoV in the proposed CEDG structure can be approximated as:

\begin{equation}
    \label{eq.24}
    FoV_{max}= 2\times(3.57\times\frac{\lambda d_s^2 F_1}{M_SD_M})^{1/3}.
\end{equation}

For example, the maximum achievable FoV for the proposed CEDG structure of Section \ref{section.4} is calculated to be 43.15 cm. Also, for the sake of comparison and observing the trade-off between these parameters, the estimated maximum obtainable FoV for $d_s$=3 m and as a function of the $D_M$ and several values of the $M_s$ are depicted in Fig. \ref{fig.7}(b). 

\begin{figure}[htbp]
\centering\includegraphics[width=11cm, trim={7cm 8cm 7cm 2cm},clip]{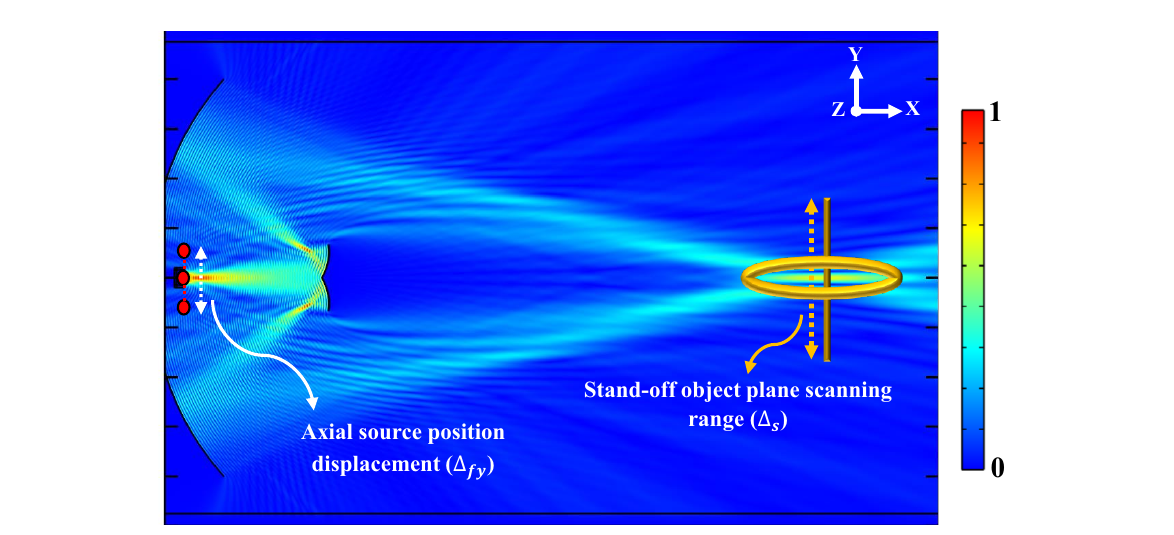}
\caption{Schematic representation of performing the stand-off object plane scanning utilizing the axial displacement of the feed position in the proposed CEDG optical configuration.}
\label{fig.8}
\end{figure}

\noindent 2- \textbf{Feed displacement}: Utilizing feed position displacement is another technique commonly used in imaging and communication systems to implement the scanning operation. As shown in Fig. \ref{fig.8}, by displacing the feed position $\Delta_{fy}$ axially along the $y$-axis in the CEDG structure, the focused beam is scanned on the stand-off focal plane by $\Delta_s$. To be more specific, the vertical movement direction of the focal spot on the object plane is inversely related to the feed position movement along the y-axis. The results of this scanning scenario for several values of feed displacement are tabulated in Table \ref{Scanning_FPA}. Based on this table, by changing the feed position in the proposed CEDG structure from 5 mm to 30 mm, the stand-off focal plane can be scanned from 7.3 cm to nearly 19.2 cm. Therefore, a FoV of nearly 40 cm is achievable, similar to the SM tilting scanning scenario. It is also worth noting that in both scanning scenarios, the maximum tolerable distortions and aberrations of the rays for the specific application determine the ultimate obtainable FoV of the imaging system.

\begin{table}[htbp]
	\centering
	\caption{\bf CEDG structure stand-off object plane scanning by the axial feed displacement.}
	\begin{tabular}{|c|c|c|c|c|c|c|}
		\hline
            \rowcolor{cyan!10}
		\textbf{$\Delta_{fy}$ [mm]} & \textbf{5} & \textbf{10} & \textbf{15}  & \textbf{20} & \textbf{25} & \textbf{30}  \\ 
		\hline
            \cellcolor{cyan!10}
		\textbf{$\Delta_s$ [cm]} & $\pm7.3$ & $\pm11.8$ & $\pm14.9$  & $\pm16.7$ & $\pm18.9$ & $\pm19.2$  \\ 
		\hline
	\end{tabular}
	\label{Scanning_FPA}
\end{table}

\subsection{Tuning the stand-off imaging distance}
\begin{figure}[htbp]
\centering\includegraphics[width=16.5cm, trim={11.2cm 18.5cm 4cm 9cm},clip]{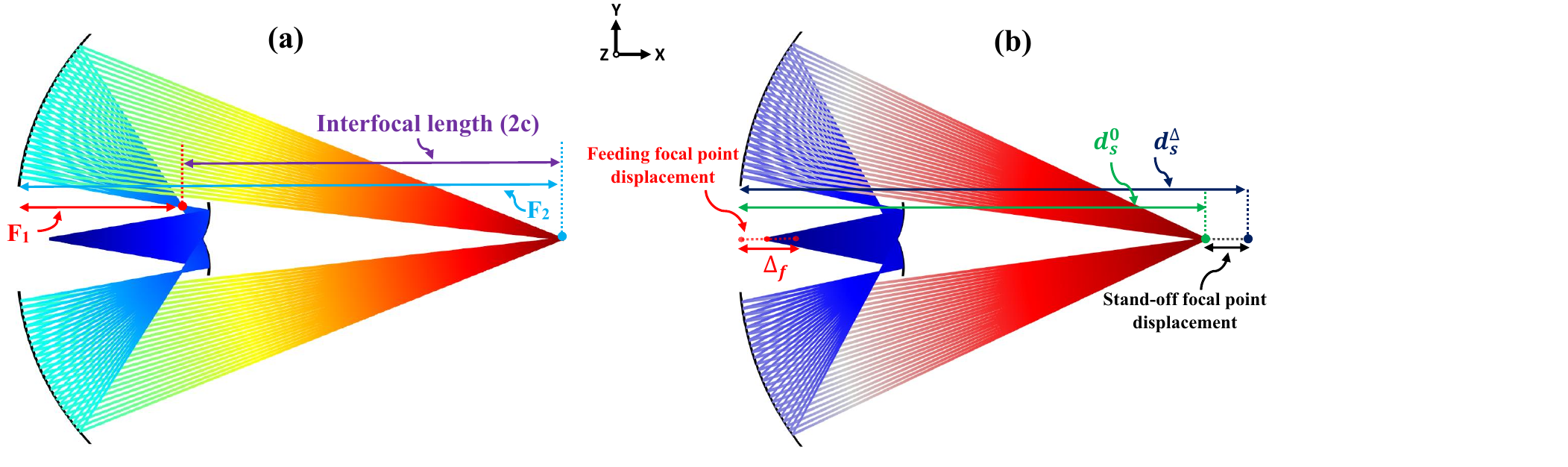}
\caption{Refocusing of the CEDG configuration by (a) manipulating the elliptical MM constructive coefficients and (b) lateral displacement of the feeding focal point.}
\label{fig.9}
\end{figure}

In certain applications, it is necessary to change the initial distance at which the object to be imaged is located. Adjusting the distance at which the beam is focused, known as refocusing, can be performed using two methods in our proposed CEDG structure:

\noindent 1- \textbf{Changing the MM foci}: The positions of the first and second focal points of the ellipsoidal MM, $F_1$ and $F_2$, can be determined as:

\begin{equation}
\left\{
\begin{array}{l}
F_1 = \frac{r}{p}(1-\sqrt{1-p}), \\
F_2 = \frac{r}{p}(1+\sqrt{1-p}). \\
\end{array}
\right.
\label{eq24}\\
\end{equation}

\noindent where $r$ and $p$ are the radius and conic constant of the MM curve respectively, and $F_2$=$F_1$+2$c$ is satisfied. Therefore, as shown in Fig. \ref{fig.9}(a), the stand-off focal point $F_2$ can be easily changed by manipulating the parameters of the elliptical MM curve.

\noindent 2- \textbf{Lateral displacement of the feed point}: Another technique for changing the ray's focusing distance is to move the emanating ray source point laterally. This method is particularly useful when the system is designed and it is not possible to change the MM curve. In this method, as depicted in Fig. \ref{fig.9}(b), the lateral displacement of the feeding focal point by $\Delta_f$ results in the stand-off focusing distance being moved from $d_s^0$ to $d_s^\Delta$. Table \ref{Refocusing} shows some examples of feed point shift and corresponding stand-off focus displacing for the three $M_s$ values of 6, 3, and 10. Additionally, the distances between the MM and SM have been chosen to avoid blockage effects when the feed point position is shifted.

\begin{table}[htbp]
	\centering
	\caption{\bf CEDG refocusing by shifting the feed position laterally.}
	\begin{tabular}{|c|c|c|c|c|c|c|c|}
		\hline
            \rowcolor{cyan!10}
		\textbf{$\Delta_{f}$ [cm]} & \textbf{0} & \textbf{-0.5} & \textbf{-1.5}  & \textbf{-2.5} & \textbf{0.5} & \textbf{1.5} & \textbf{3.5} \\ 
		\hline
            \cellcolor{cyan!10}
		\textbf{$d_s^{\Delta}$ [m] ($M_S$=6)} & 3 & 3.3 & 4.3 & 6.2 & 2.7 & 2.3 & 1.8  \\ 
		\hline
            \cellcolor{cyan!10}
         \textbf{$d_s^{\Delta}$ [m] ($M_S$=3)} & 3 & 3.4 & 4.5 & 6.6 & 2.5 & 1.8 & 1.1  \\ 
		\hline
            \cellcolor{cyan!10}
           \textbf{$d_s^{\Delta}$ [m] ($M_S$=10)} & 3 & 3.1 & 3.4 & 3.7 & 2.8 & 2.6 & 2.4  \\ 
		\hline
	\end{tabular}
	\label{Refocusing}
\end{table}

Based on the results of Table \ref{Refocusing}, the lateral displacement of the feed position along the optical axis can significantly change the stand-off imaging distance, which can be a favorable option in practical implementations. Furthermore, the amount of stand-off distance change is inversely related to the optical magnification of the mirror setup. To be more specific, increasing $M_s$ reduces the rate of stand-off distance change, and vice versa. For example, for the CEDG system designed for a stand-off distance of $d_s^0$=3 m with $M_s$=6, displacing the feed position by ${\Delta_f}$=+3.5 cm on the optical axis towards the focal plane causes the rays to refocus at the distance of $d_s^{\Delta}$=1.8 m. Also, in the case of $M_s$=3 and $M_s$=10, the stand-off distance is reduced in the range of $d_s^{\Delta}$=1.1 m and $d_s^{\Delta}$=2.4 m, respectively. On the other hand, the focusing distance can be increased to $d_s^{\Delta}$=6.6 m with just about ${\Delta}_f$=-2.5 cm of lateral feed displacement in the reverse direction of the optical axis for the magnification value of $M_s$=3.

\section{Conclusion} \label{section.6}
In this paper, a blockage-free confocal ellipsoidal displaced Gregorian (CEDG) scannable structure is proposed for stand-off imaging applications. A straightforward design procedure is presented and the analytical formulas are derived for constructing the structure. The characteristics of the proposed configuration are investigated by a design example, and the ray tracing results are compared with the classical Gregorian configuration, showing considerable improvement in terms of the blocked rays. Furthermore, the ability of the proposed CEDG configuration to be used as a near-field scannable imaging structure is proved in terms of achievable resolution, DoF, FoV, and stand-off focusing distance tunability using a combination of full-wave and ray tracing simulations. As a result, the proposed CEDG reflective structure can be an attractive candidate for use in the optical section of the mm-wave and THz stand-off imaging systems.


\section*{Disclosures}
The authors declare that there are no conflicts of interest
related to this article.

\section*{Data Availability Statement}
Data underlying the results presented in this paper are not publicly available at this time but may be obtained from the authors upon reasonable request.

\section*{Supplemental document}
See Supplement 1 for supporting content.


\bibliographystyle{unsrt}
\bibliography{references}






\end{document}